# Restricted Hilbert transform for non-Hermitian management of fields

W. W. Ahmed[1], R. Herrero[2], M. Botey[2], Y. Wu[1*], and K. Staliunas[2,3*]

[1]*Division of Computer, Electrical and Mathematical Sciences and Engineering, King Abdullah University of Science and Technology (KAUST), Thuwal, 23955-6900, Saudi Arabia*

[2]*Departament de Física, Universitat Politècnica de Catalunya (UPC), Colom 11, E-08222 Terrassa, Barcelona, Spain*

[3]*Institució Catalana de Recercai Estudis Avançats (ICREA), Passeig Lluís Companys 23, E-08010, Barcelona, Spain*

Emails: ying.wu@kaust.edu.sa, kestutis.staliunas@icrea.cat

**Abstract:** Non-Hermitian systems exploiting the synergy between gain and loss have recently become the focus of interest to discover novel physical phenomena. The spatial symmetry breaking in such systems allows tailoring the wave propagation at will. Inspired by such property, we propose a feasible approach based on local Hilbert transform to control the field flows in two- or higher dimensional non-Hermitian systems, restricting the complex refractive index within practical limits. We propose an iterative procedure to reduce the dimensionality of complex refractive index parameter space to two, one or zero dimensions. The proposed method provides a flexible way to systematically design locally PT-symmetric systems realizable with a limited collection of realistic materials.

**Introduction:** The celebrated Hilbert Transform (HT), is known to be closely related to causality. One representative example of HT in optics is the Kramers-Kronig (KK) relations, implying that the response of a system must occur at a later (not earlier) time of an excitation. It relates the real and imaginary parts of the spectrum of the response function[1]. Such a HT breaks the time symmetry, which accounts for the "*invisibility*" of the future.

Recently, an analogous relation was considered in space. The spatial profile of the complex susceptibility of a system can be engineered to non-Hermitically manipulate the direction of the response in space, breaking the spatial symmetry. In one-dimension (1D), the space and time are both scalars and are mathematically equivalent for classical waves[2]. The invisibility of the future or the absence of the flow of information to the past corresponds to the absence of scattering, say backscattering irradiating from the direction opposite to the illumination[3-5]. In

two- and three-dimensions, however, the situation becomes more complicated as backscattering (unidirectional invisibility) can be eliminated from different directions. For instance, in a two-dimensional (2D) space, it is possible to manipulate the response (scattering) of an object toward particular directions[6,7] by correspondingly modifying its spatial response function (susceptibility) via HT.

The spatial symmetry breaking of the system response, on the global scale is extensively studied in the context of Parity-Time (PT)-symmetry to explore novel physical effects[8-18]. Moreover, such symmetry breaking of the spatial response may occur locally, i.e. distinct at different spatial positions, which allows engineering of complicated directionality fields, with particular topologies such as axisymmetric[19] or with arbitrary configurations being from periodic, quasiperiodic, to random[20].

Despite some attempts[7,21], a practical realization of HT in space, either on a global or local scale, is still a grand challenge, because the possibilities of the realization of spatial HT are severely restricted by the limited availability of materials that can satisfying the required complex refractive index profile ($n_{Re}, n_{Im}$). With the development of metamaterials, which derive their response function from their subwavelength artificial structures rather than chemical compositions, the possible regime of refractive index profile has been greatly expanded, bringing possibilities to realize a precise relation between the real/imaginary parts of the response function required by the spatial HT. For instance, metamaterials based on a particular collection of "metachips" satisfying a given response spectra can realize a spatial HT profile in the microwave regime[22]. However, it is not possible to realize arbitrary relations between real/imaginary parts of susceptibility from a–limited collection of real materials. In semiconductor microstructures, the situation is also similar, as actual active materials have a specific relation between index and gain-loss.

The fundamental question that arises is whether it is possible to implement an actual field of unidirectionality, either with a global or local topology, using the HT approach, using restricted (realistic) range of complex susceptibility of a material. Arbitrary unidirectional vector fields can be designed by locally modifying the scattering using the local Hilbert transform. As an example, Fig.1 illustrates such directionality field consisting of a sink and a source. However, the resultant complex index profile demands a large number of actual materials for the practical realization due to the continuous variation of complex index values in space [see Fig. 1(b) and 1(c)]. To resolve this issue, we propose to introduce restrictions for the complex index values, constraining them around certain particular values or confining them in some particular areas of the parameter space $(n_{Re}, n_{Im})$. Such a restriction may possibly not limit the field unidirectionality or total invisibility, resulting into a basic scattering reduction. This article proposes an iterative procedure, to include such constraints to design the locally PT-symmetric systems with realistic index and gain/loss values while keeping a significant field unidirectionality.

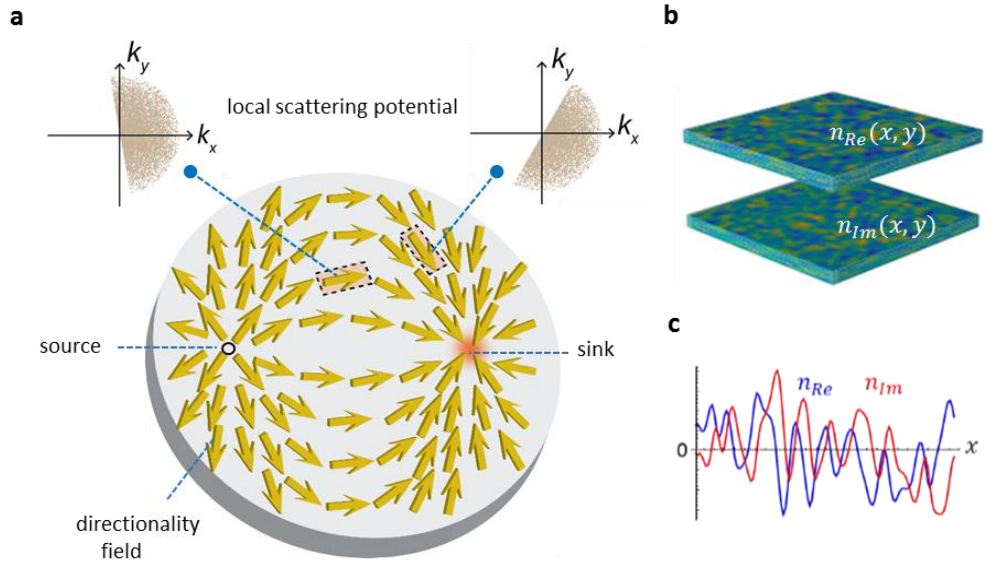

Figure 1 | Locally modified Hilbert transform media (a) Scattering potential of an optical media, $n_{Re}(x, y)$, modified by local HT to mold the flow of light in the desired direction. Here, the optical media is specifically tailored to create source and sink fields, which involves gain-loss regions, $n_{Im}(x, y)$ (b) 2D complex refractive index profiles of the modified HT media requiring a large number of materials corresponding to different spatial point (c) cross-sectional profile of the complex index.

The different restrictions of the non-Hermitian media following the HT considered in our study are the following:

(1) Preparation of the HT media by using a mixture of two different materials with two different complex refraction indices. Restricting the total filling factor of two materials leads to a 1D manifold in the refractive index complex parameter space (n-complex space), while limiting the total filling factor of two materials to some upper bound results in a general 2D manifold of complex refraction indices, a surface [Fig. 2(a)]. Mixing of chemical components or/and manipulating porosity is a common practice to vary the refraction indices in desired range typically between 1.2 to 2.1 in vapor deposition techniques[23,24].

(2) Preparation of the HT media from metamaterials consisting of continuous family of size-scaled metachips in microwave range[22], which restricts the complex refraction index space to a ring [Fig. 2(b)].

(3) Preparation of the HT media with a "poor man" collection of the size-scaled resonators (for instance, of Helmholtz resonators in acoustics, or microwave metachips), corresponding to a discrete set of points in the complex index space [Fig. 2(c)].

In this way, to show the universality of the method to engineer the desired HT profiles in the restricted parameter space, we analyze these distinct cases, which entail reducing the dimensionality in parameter space. For instance, Fig. 2 illustrates the kind of the dimensionality reductions of the three proposed restrictions (a) 2D→2D, (b) 2D→1D and (c) 2D→0D.

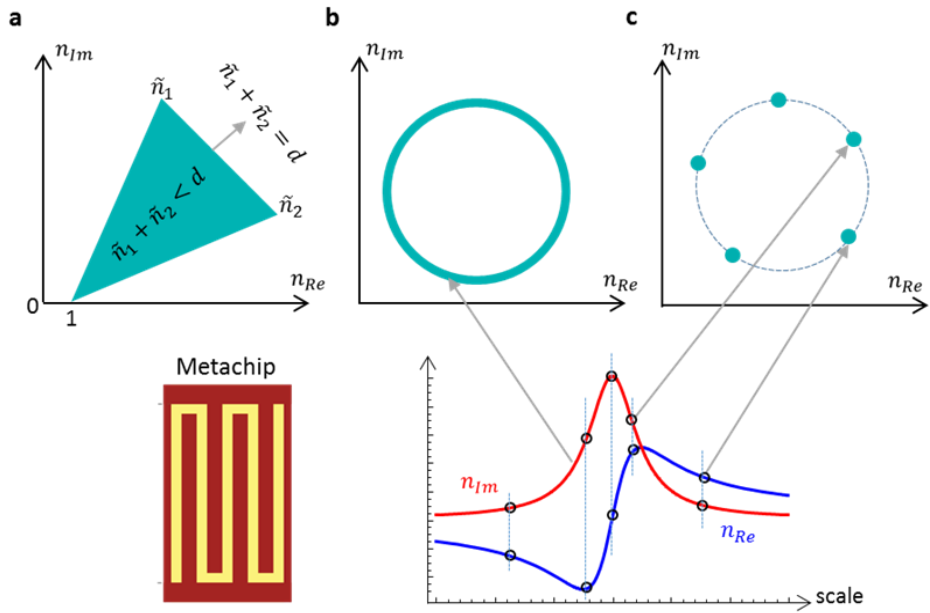

Figure 2 | Restrictions of the local HT for different dimensionalities (a) mixture of two different materials with $\tilde{n}_1$, $\tilde{n}_2$; for instance restriction of constant total density $\tilde{n}_1 + \tilde{n}_2 = d$ (solid or liquid –like) gives an available line, and the mixture of two materials with densities $\tilde{n}_1 + \tilde{n}_2 < d$, gives a 2D area in complex space, (b) continuum of scalable in size metachips gives a ring in complex refractive index space, and (c) limited collection of scaled-in-size meta-atoms gives a set of points in complex space (five points in this case).

Besides, there is still another fundamental aspect: the ideal HT requires the modification of the material parameters over all space (as for KK in time, where causality requires the modification of all spectrum, although the term $1/(\omega - \omega_{res})$ rapidly decays away from the resonance frequencies). Restricting either the index or the gain-loss range in the *n*-complex space may lead to unrealistic divergent values of the corresponding real or imaginary values, because the spatial derivatives of the parameters become large when the values are restricted to small spatial areas. This imposes the question again, whether the restricted HT helps to remain within the possible range of real/imaginary values.

**Results:**

**Restricted Local Hilbert transform**: To implement the proposed schemes, we apply an iterative procedure. First, we perform a local HT on an arbitrary background potential to create intended desired directionality flows such as a sink, which requires a large area of real/imaginary values of $n(\vec{r})$. Second, we move the unallowed values to the restriction area. This process "spoils" the HT. Finally, we perform the HT again and repeat the procedure until the index values converge. We define a correlation coefficient for the generated vectorial potentials with different restriction dimensionalities to characterize the accuracy and convergence of the restricted HT, which confirm that this iterative approach leads to the converging results in all cases studied. To verify the directionality effect in restricted HT, we perform numerical simulations using the Schrödinger equation (for paraxial optics, or zero-temperature Bose condensates) for linear systems with given complex vectorial potentials. We also perform the full wave simulations to demonstrate the functionality of the proposal beyond the paraxial approximation (see Supplementary 1.3).

As an example of a possible realization of a restricted HT media, we consider an initial real-valued hexagonal profile, $n_{Re}(\vec{r})$ [see top-left panel in Fig. 3] and generate the corresponding gain-loss profile, $n_{Im}(\vec{r})$ by applying the local HT transform[20] to ensure the sink directionality: $p(\vec{r}) = -\vec{r}/|\vec{r}|$. The parameter space of complex refractive index $(n_{Re}, n_{Im})$ provides different material parameters as functions of the spatial location [see Fig. 3(a)]. To realize such complex profiles with available materials, we apply local HT iteratively to restrain the complex index values within physical limits. The results for different restriction dimensionalities are shown in Fig. 3(b-d). The index values in complex space $(n_{Re}, n_{Im})$ illustrate that the iterative procedure precisely limits the refractive indices within the designated area, ring or a set of points on a ring [see the third row of Fig. 3]. The corresponding density distributions of complex index, plotted in the last row of Fig. 3, also show the spreading of restricted values inside the desired bounds. Note that the restricted HT provides many possible ways to restrict the complex indices during the iterative procedure and some of them are discussed in Supplementary Note 1. In addition, the procedure is independent of the background potential profile and can be applied to arbitrary initial distribution i.e. quasiperiodic, random, localized etc [see Supplementary Note 2 for random background medium].

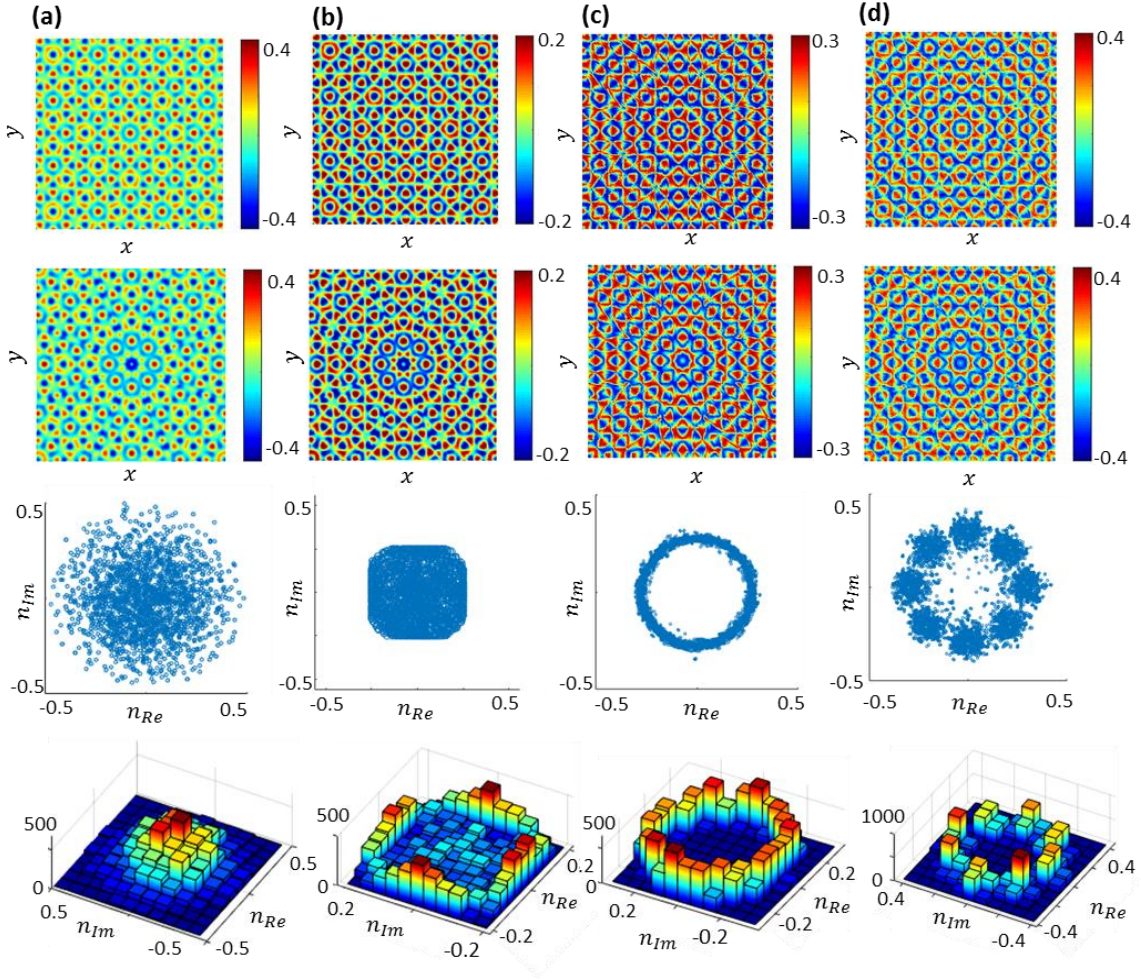

Figure 3 | Complex refractive index distributions for sink directionality (a) local HT (b,c,d) Restricted local HT (b,c,d) with hexagonal background pattern. In restricted HT cases, the complex index profiles are obtained after fifteen iterations. The first row presents the real part of refractive index, second row depicts the imaginary part obtained from the local HT, and third and fourth rows depict the corresponding restricted index values and density distribution in complex space, respectively.

**Convergence of Restricted Hilbert transform:** To analyze the robustness and convergence of the iterative procedure, we determine the correlation coefficients in terms of the flow of the complex potential generated by local HT under different restrictions. We define the correlation coefficient as $C = \int p(\vec{r}) \cdot f_k(\vec{r}) dr / \sqrt{\int |p(\vec{r})|^2 dr^2 \cdot \int |f_k(\vec{r})|^2 dr^2}$ where $p(\vec{r})$ is the reference flow and $f_k(\vec{r}) = i(U_k \nabla U_k^* - U_k^* \nabla U_k)$ is the potential flow after $k$ iterations with the restrictions. For the cases shown in Fig. 3, we assume sink directionality to generate the complex vectorial potentials. Therefore, the reference flow is kept with the form of the sink: $p(\vec{r}) = -\vec{r}/|\vec{r}|$ to determine the correlation coefficient. We expect that the constraints have a weak influence on the pattern of potential flows with reference to sink. Here, we compute two correlation coefficients namely $C_0$ and $C_1$ for which $p_k(\vec{r})$ is calculated from complex refractive distribution obtained by applying HT before and after the restriction at $k$ iteration, respectively.

The difference between $C_0$ and $C_1$ estimates the accuracy of the proposed restricted HT. To illustrate the difference of applying the HT before and after restrictions, we present the refractive indices in complex space after 1st and 12th iterations in Figs. 4(a-c). The second row in Figs. 4(a-c) shows the converged values of complex refractive indices for all cases. In (a), the restricted HT accurately limit the complex refractive indices in desired ranges. However, for (b,c), the refractive indices somehow spread around the ring and the chosen discrete points. This spreading behavior due to restrictions in complex space may be associated with uncertainty principle i.e. the smaller the restricted area, the larger the spreading. The correlation coefficients $C_0$ and $C_1$, calculated during the iterative procedure, are plotted in Fig. 4(d). In the square case, we obtain exactly the same values for both correlation coefficients after fifteen iterations ensuring 100% accuracy. However, we found ~ 1% and ~ 5% difference in correlation coefficients for ring and discrete cases, respectively.

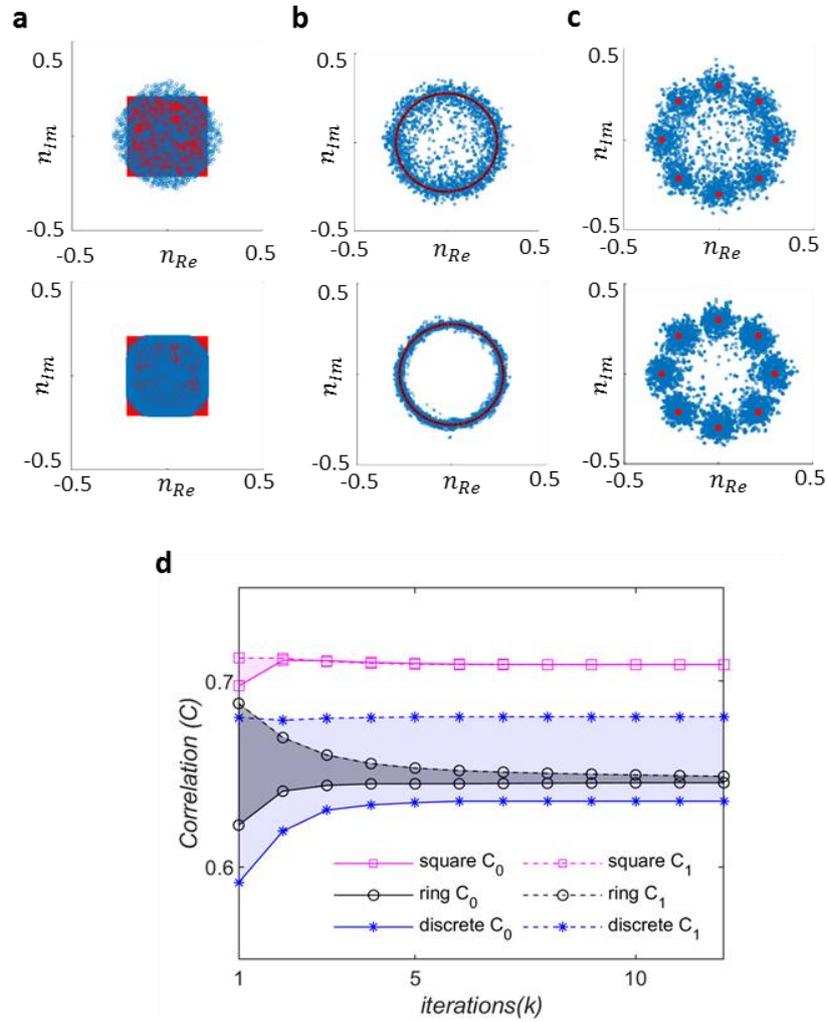

Figure 4 | Convergence of restricted local Hilbert transform. (a-c) first and second rows depict refractive indices in complex space after 1st and 12th iterations, respectively. The red and blue color illustrate the desired and computed restricted complex refractive index values obtained from local HT, respectively. (d) Correlation

coefficients ($C_0$ solid line ; $C_1$ dotted line) in terms of flow of the complex potential generated by HT during iterative procedure for different restriction dimensionalities (□) 2 D→2D, (○)2D→1D, (*) 2D→0D.

Note that the difference between correlation coefficients, for the considered restrictions, remains the same regardless of background pattern but the exact correlation values, $C_0$ and $C_1$, depend on the background pattern. For instance, in pure sinusoidal pattern, these values are close to one but lower than unity in the considered hexagonal pattern. We note that the convergence of correlation coefficients requires more iterations when the restriction area shrinks. In the square case, the large restriction area leads to a fast converge, as compared to ring and discrete case. However, the iterative procedure converges after ten iterations in all cases as depicted in Fig. 4(d).

**Field evolution in Restricted HT media**: To validate our proposed "*Restricted Hilbert transform*", we perform numerical simulations using paraxial equation of diffraction (mathematically equivalent to the Schrodinger equation for a quantum wave function) expressed in the form:

$$\partial_t A(\vec{r},t) = i\nabla_\perp^2 A(\vec{r},t) + iU(\vec{r})A(\vec{r},t) \qquad (1)$$

where $A(\vec{r},t)$ is a slowly varying complex field envelop distributed in space, $\vec{r} = (x,y)$ and evolving in time, $t$. $U(\vec{r}) = n_{Re}(\vec{r}) + in_{Im}(\vec{r})$ is the non-Hermintian potential being $n_{Re}(\vec{r})$ the real refractive index profile and $n_{Im}(\vec{r})$ the corresponding imaginary part of the potential obtained from local HT with different restrictions. We numerically solve Eq. (1) using the split step method for HT media with different restrictions. For the complex refractive index profiles shown in Figs. 3(a–d), the simulated steady-state field distributions are provided in Fig. 5, where a gaussian source is initially placed at an arbitrary position within the modified structure as show in Fig. 5a(i). The system initially shows transients as depicted in Fig. 5a(ii) but eventually the field is efficiently concentrated around the center in a(iii), due to the sink directionality. The final localized states for different restrictions are shown in (b-d).

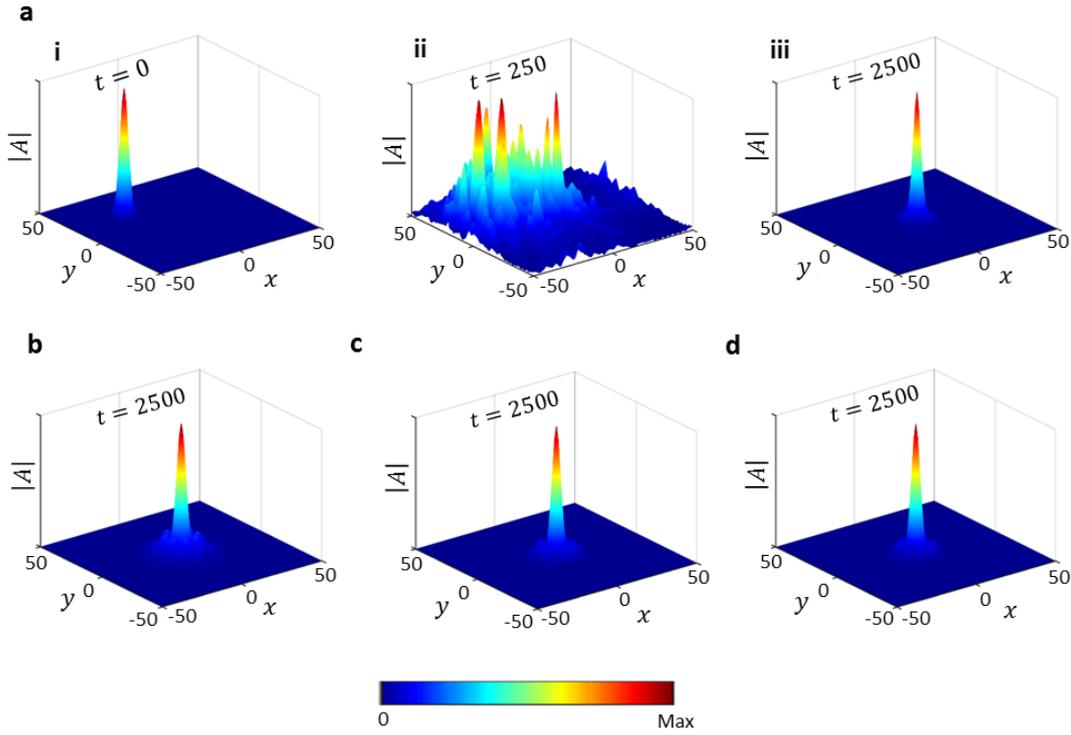

Figure 5 | Field evolution in modified HT media for sink directionality. The system is excited with a gaussian beam placed at an arbitrary position within the structure as shown in a(i). Numerically calculated transient and final localized states are shown in a(ii) and a(iii), respectively. The steady state distributions for restricted cases are: (b) square area (c) ring (d) a set of discrete points on the ring. The results indicate that gaussian source, initially located at arbitrary position, localizes around the center in all cases after sufficient propagation time.

To conclude, we propose a restricted Hilbert transform for the generation of feasible systems holding local PT-symmetry to manage arbitrary field flows in higher dimensional non-Hermitian systems under a realistic parameter space. In particular, we study restrictions of different dimensionalities from 2D to 1D and 0D to achieve the refractive indices for locally PT-symmetric systems within desired ranges. The procedure provides a substantial control over the chosen index values as compared to conventional spatial Kramers-Kronig relation, and the constructed index profiles can be experimentally realized with limited collection of realistic materials by locally tuning the real and imaginary part of dielectric media. It also offers a general design strategy to implement any desired field configuration in a broad class of non-Hermitian systems conceptually different from existing coordinate transformation approaches[25-28]. We believe that the proposal opens new possibilities to practically realize the wave dynamics of linear and nonlinear physical systems based on engineered HT media with realistic index and gain/loss values.

## Methods

**Theory:** Local Hilbert transform is employed to calculate the non-Hermitian potentials with the desired directionality from an arbitrary background pattern. It relates the two quadratures of a complex potential that are the real, $n_{Re}(\vec{r})$, and imaginary, $n_{Im}(\vec{r})$, parts of the potential, corresponding to the refraction index and gain-loss in optics. In two-dimensional systems, the pair of local Hilbert transform is expressed as:

$$n_{Re}(\vec{r}) = \frac{1}{\pi} P.V. \iint \frac{\delta((\vec{r}-\vec{r}_1)\cdot\vec{q}(\vec{r}-\vec{r}_1))n(\vec{r}_1)}{\vec{p}(\vec{r})(\vec{r}-\vec{r}_1)} d\vec{r}_1 \; ;$$

$$n_{Im}(\vec{r}) = \frac{-1}{\pi} P.V. \iint \frac{\delta((\vec{r}-\vec{r}_1)\cdot\vec{q}(\vec{r}-\vec{r}_1))n(\vec{r}_1)}{\vec{p}(\vec{r}_1)(\vec{r}-\vec{r}_1)} d\vec{r}_1 \quad (2)$$

where $\delta$ is the Kronecker-delta function, *P.V.* is principal value of the integral, $\vec{p}(\vec{r})$ is the directionality of scattering and $\vec{q}(\vec{r})$ is a unit vectorial field orthogonal to a given field of directionality, $\vec{q}(\vec{r}) \cdot \vec{p}(\vec{r}) = 0$.

**Numerical Simulations:** The numerical simulations have been performed by solving the Schrödinger equation for the complex non-Hermitian potential, obtained from Eq.(2), using split step method. An initial broad Gaussian beam is considered at an arbitrary position in a 2D computation window of size $100 \times 100$ to record the evolution of field in HT media under different proposed restrictions.

**Acknowledgements:** The work described in here is partially supported by King Abdullah University of Science and Technology (KAUST) Office of Sponsored Research (OSR) under Award No. OSR-2016-CRG5-2950 and KAUST Baseline Research Fund BAS/1/1626-01-01. K. Staliunas acknowledges the support of Spanish Ministerio de Economía y Competitividad (FIS2015-65998-C2-1-P) and European Union Horizon 2020 Framework EUROSTARS (E10524 HIP-Laser).

**Author contributions:** K. S, R. H and M. B proposed the idea. W.W. A conducted the analysis and simulations. K. S. and Y. W. supervised the project. All authors were involved in the discussions on the idea and methods, on writing and revising the manuscript.

# Supplementary Information:

In this supplementary material, we discuss different specific restriction strategies in the proposed procedure to implement the restricted Hilbert Transform (HT), and provide a wider class of examples to support our theory. We demonstrate that the proposal also applies for random systems and can be realized beyond the paraxial limit by performing full wave time-domain simulations using COMSOL.

**Supplementary Note 1: Different restriction procedures in Restricted Hilbert transform**

The proposed theory offers many possible ways to restrict the index values in desired complex space during the iterative procedure. Here, we discuss different restriction strategies to achieve the index values in the desired ranges. For instance we can restrict the index so that all values lay within some specified area in the complex refractive index plane, for instance within a ring with given thickness. To achieve this, we may follow different approaches by randomly redistributing the unallowed values to the central circle or to the edges of such ring area. The results for this square case are shown in Fig. S1.

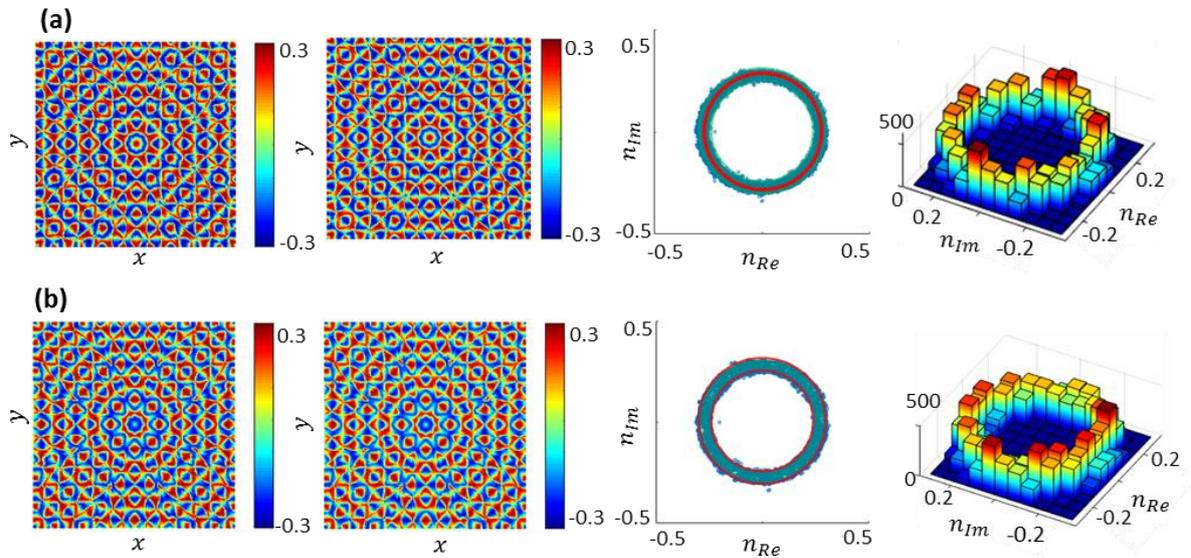

Fig. S1. HT media with octagonal background symmetry, imposing a ring-shaped restriction area (in green) by iteratively randomly distributing the unallowed values to: (a) the circle at the center of the ring, or to (b) the edges of the ring, as indicted by red lines color in the panels of the third column. First column illustrates the real part of the index, second column the imaginary part, third column the distribution of index values in complex space (in blue), and fourth column density distribution of the corresponding index values in complex space parameter.

We may also consider the restriction in the form of small circular disks by redistributing the unallowed index values, external to the disks area, either to the center of the disks or edges of the disks during the iterative procedure. The results for this circular disks case are shown in Fig.

S2. Note that according to the procedure followed, the final two distributions are more different in this case, compare the fourth columns on Fig. S1.

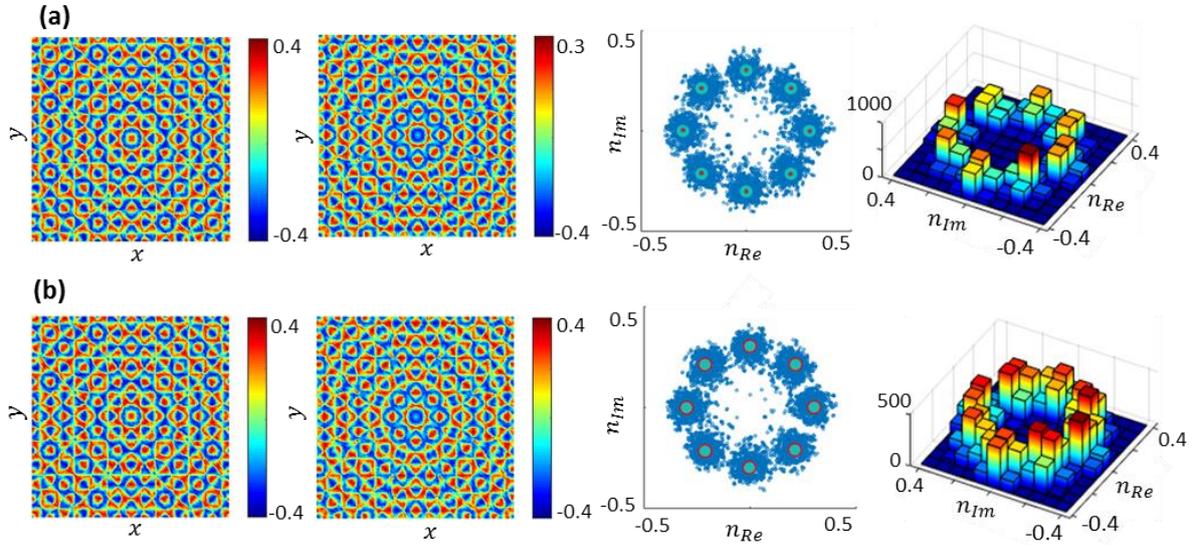

Fig. S2. HT media with octagonal background symmetry, imposing discrete disk-shaped restriction area (in green) by iteratively randomly distributing the unallowed values to: (a) the center of the disks, or to (b) the edges of the disks, as indicted by red color in the panels of the third column. First column illustrates the real part of the index, second column the imaginary part, third column the distribution of index values in complex space and fourth column density distribution of the corresponding index values in complex space parameter.

Moreover, we consider the restriction in the form of a square area by randomly moving the unallowed values, external to the square area, either to a square framed shape or the whole squared area, during the iterative procedure. The results are shown in Fig. S3.

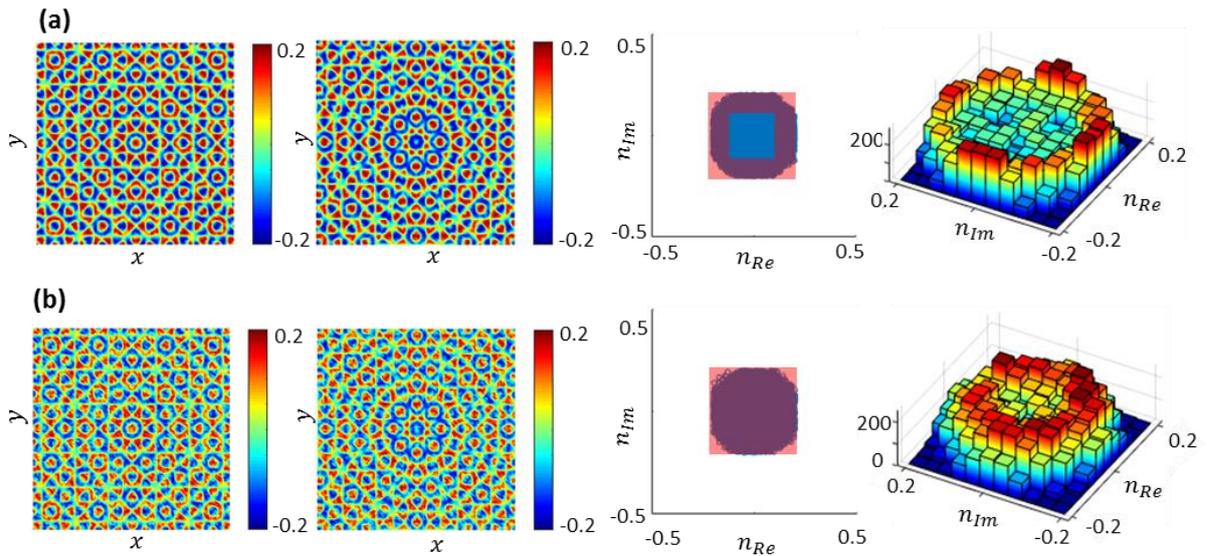

Fig. S3. HT media with octagonal background symmetry, imposing a square-shaped restriction area by iteratively randomly distributing the unallowed values to: (a) a specified square-framed ring area (b) within the whole square. First column illustrates

the real part of the index, second column the imaginary part, third column the distribution of index values in complex space (in blue), and fourth column density distribution of the corresponding index values in complex space parameter.

**Supplementary Note 2: Restricted Hilbert transform for Random media**

To show the realization of restricted HT in random system, we consider a real-valued random pattern and determine the corresponding gain-loss profile using HT iterative procedure for sink directionality. The results for different restriction, same as in the main manuscript, are presented in Fig. S4. The numerical simulations show that the fields are efficiently accumulated around the center, in all cases, due the sink behavior [see the last row of Fig. S4].

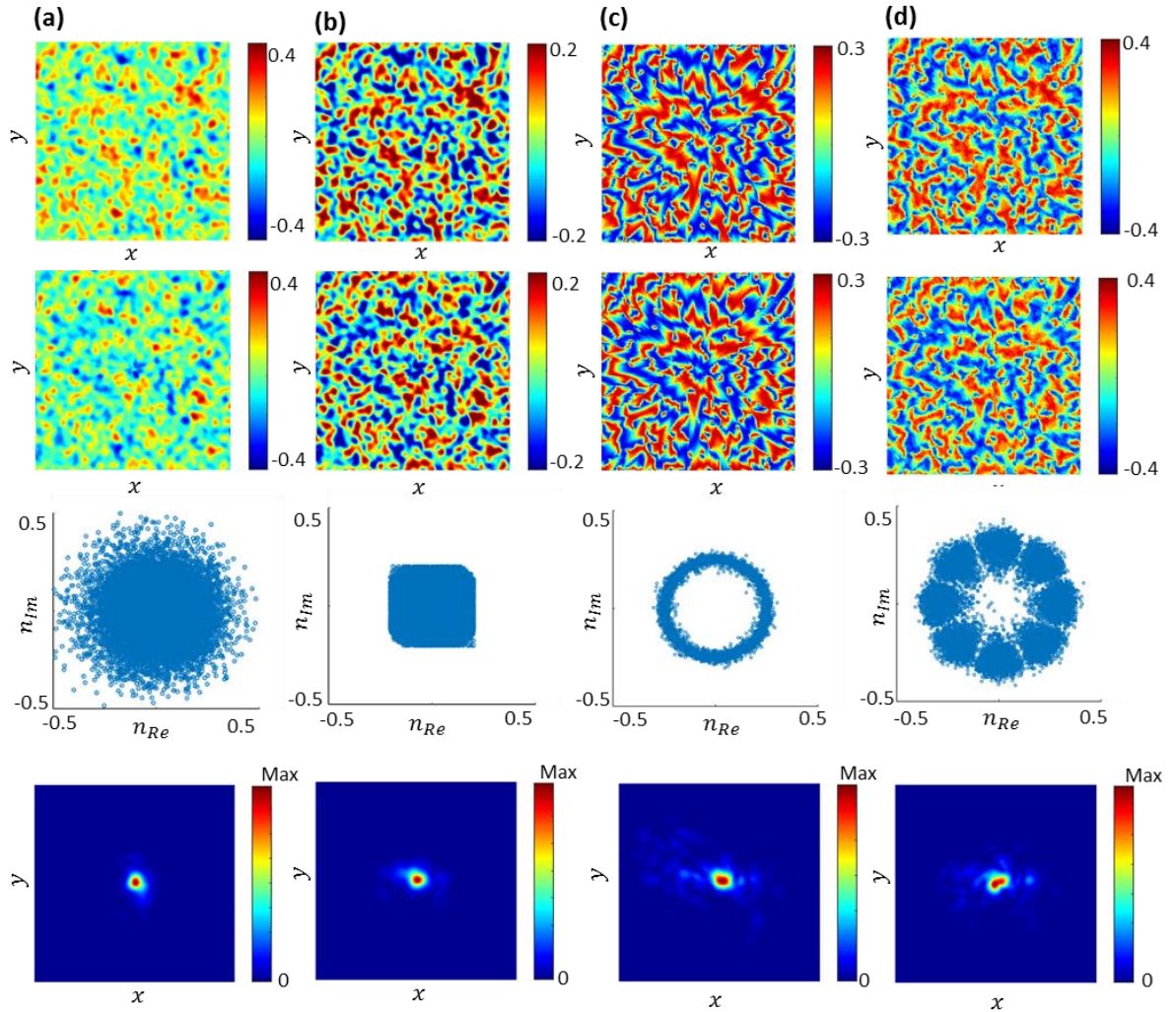

Fig. S4. Random background pattern (a) Unrestricted HT, (b) Restricted HT media showing refractive indices within desired square area (c) ring (d) a set of discrete points on the HT ring. First row illustrates the real part of the index, second row the imaginary part, third row the distribution of index values in complex space and forth row shows the numerically calculated steady state field distributions using Schrodinger model. The probe field shows intense localization around the center for a Gaussian beam source placed at an arbitray position within the structure.

**Supplementary Note 3: Full wave simulations for Restricted Hilbert transform**

To verify the "*Restricted Hilbert transform*" beyond the paraxial approximation, we perform full wave time domain simulations using COMSOL Multiphysics, by placing a point source within the HT media. The complex refractive index profiles for different restriction are provided in the first and second row of the Fig. S5. The point source radiating electric field polarized perpendicularly to the plane, is located at $(x,y) = (-10\lambda_0, 10\lambda_0)$. In Fig. S5, the bottom row shows the steady state field profiles calculated after sufficient propagation time. The results clearly indicate that the vectorial potential concentrates the field at the center, due sink directionality, in all cases. Note that the point source becomes imperceptible after certain propagation time due to the amplification of the field around the center in the linear system.

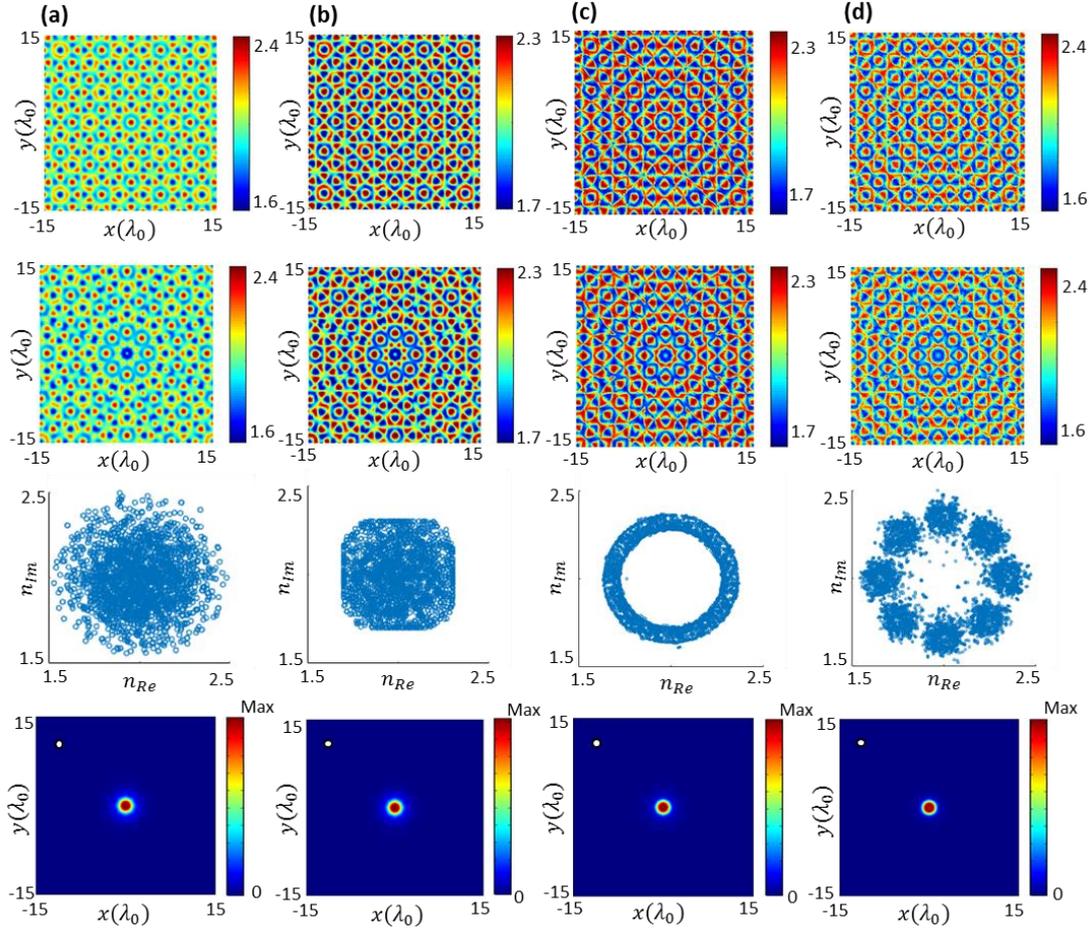

Fig. S5. Hexagonal background pattern (a) Unrestricted HT, (b) Restricted HT media showing refractive indices within desired square area (c) ring (d) a set of discrete points on the ring. First row illustrates the real part of the index, second row the imaginary part, third row the distribution of index values in complex space and forth row shows the numerically calculated steady state field distributions using COMSOL. The point source, located at $(x,y) = (-10\lambda_0, 10\lambda_0)$, radiates energy and eventually the probe field is confined around the center.

# References

1. Jackson, J. D. Classical Electrodynamics (Wiley, 1999).
2. Horsley S. A. R., Artoni M. & La Rocca G. C. Spatial Kramers-Kronig relations and the reflection of waves. *Nat. Photon*. 9, 436–439 (2015).
3. Longhi, S. Half-spectral unidirectional invisibility in non-Hermitian periodic optical structures, *Opt. Lett*. 40, 5694-5697 (2015).
4. Longhi, S. Wave reflection in dielectric media obeying spatial Kramers-Kronig relations. *Europhys. Lett*. 112, 64001 (2015).
5. Horsley, S. A. R., King, C. G. & Philbin, T. G. Wave propagation in complex coordinates. *J. Opt*. 18, 044016 (2016).
6. Hayran, Z., Herrero, R., Botey, M., Kurt, H. & Staliunas, K. Invisibility on demand based on a generalized Hilbert transform. *Phys. Rev. A* 98, 013822 (2018).
7. Hayran, Z., Herrero, R., Botey, M., Kurt, H. & Staliunas, K. All-Dielectric Self-Cloaked Structures. *ACS Photonics* 5, 2068-2073 (2018).
8. Makris, K. G., El-Ganainy, R., Christodoulides, D. N. & Musslimani, Z. H. Beam dynamics in PT symmetric optical lattices. *Phys. Rev. Lett*. 100, 103904 (2008).
9. Guo, A. et al. Observation of PT-symmetry breaking in complex optical potentials. *Phys. Rev. Lett*. 103, 093902 (2009).
10. Rüter, C. E. et al. Observation of parity–time symmetry in optics. *Nature Phys*. 6, 192–195 (2010).
11. Longhi, S. Bloch oscillations in complex crystals with PT symmetry. *Phys. Rev. Lett*. 103, 123601 (2009).
12. Chong, Y. D., Ge, L. & Stone, A. D. PT-symmetry breaking and laser absorber modes in optical scattering systems. *Phys. Rev. Lett*. 106, 093902 (2011).
13. Lin, Z. et al. Unidirectional invisibility induced by PT-symmetric periodic structures. *Phys. Rev. Lett*. 106, 213901 (2011).
14. Feng, L. et al. Experimental demonstration of a unidirectional reflectionless parity–time metamaterial at optical frequencies. *Nature Mater*. 12, 108–113 (2012).
15. Regensburger, A. et al. Parity–time synthetic photonic lattices. *Nature* 488, 167–171 (2012).
16. Peng B. et al. Parity–time-symmetric whispering-gallery microcavities. *Nature Phys.* 10, 394–398 (2014).

17. Makris, K. G., Brandstötter, A., Ambichl, P., Musslimani, Z. H. & Rotter, S. Wave propagation through disordered media without backscattering and intensity variations. *Light Sci. Appl*. 6, e17035 (2017).
18. El-Ganainy, R., Makris, K. G., Khajavikhan, M., Musslimani, Z. H., Rotter, S. & Christodoulides, D. N. *Nature Phys*. 14, 11-19 (2018).
19. Ahmed, W. W., Herrero, R., Botey, M. & Staliunas, K. Locally parity-time-symmetric and globally parity-symmetric systems. *Phys. Rev. A*. 94, 053819 (2016).
20. Ahmed, W. W., Herrero, R., Botey, M., Hayran, Z., Kurt, H. & Staliunas, K. Directionality fields generated by a local Hilbert transform. *Phys. Rev. A* 97, 033824 (2018).
21. Jiang, W., Ma, Y., Juan, Y., Yin, G., Wu, W. & He, S. Deformable broadband metamaterial absorbers engineered with an analytical spatial Kramers-Kronig permittivity profile. *Laser Photon. Rev*. 11, 1600253 (2017).
22. Ye, D., Cao, C., Zhou, T., Huangpu, J., Zheng, G. & Ran, L. Observation of reflectionless absorption due to spatial Kramers–Kronig profile. *Nat. Comm*. 8, (2017).
23. Thelen, A. Design of optical interference coatings (McGraw-Hill, New York, 1989).
24. Tolenis, T., Grinevičiūtė, L., Buzelis, R., Smalakys, L., Pupka, E., Melnikas, S., Selskis, A., Drazdys, R., & Melninkaitis, A. Sculptured anti-reflection coatings for high power lasers, *Opt. Mater. Express* 7, 1249-1258 (2017).
25. Pendry, J. B., Schurig, D. & Smith, D. R. Controlling Electromagnetic Fields. *Science* 312, 1780-1782 (2006).
26. Leonhardt, U. Optical Conformal Mapping. *Science* 312, 1777-1780 (2006).
27. Chen, H., Chen, C. T. & Sheng, P. Transformation optics and metamaterials. *Nature Mater*. 9, 387-396 (2010).
28. Xu, L. and H. Chen, Conformal transformation optics. *Nature Photon*. 9, 15-23 (2015).